\documentclass[twocolumn]{aastex63}

\usepackage{makecell}

\accepted{02.02.2022}
\submitjournal{ApJ}

\shorttitle{\NameOfCode{}}
\shortauthors{Keil et al.}
\graphicspath{{./}{figures/}}
\begin{document}

\title{\NameOfCode{}: A MCMC Inference tool for Physical Parameters of Molecular Clouds}

\email{marcus.keil.19@ucl.ac.uk}

\author{Marcus Keil}
\affiliation{Univsersity College London, Gower St, Bloomsbury, London WC1E 6BT, England}
\author{Serena Viti}
\affiliation{Leiden Observatory, Leiden University, P.O. Box 9513, 2300 RA Leiden, The Netherlands}
\affiliation{Univsersity College London, Gower St, Bloomsbury, London WC1E 6BT, England}
\author{Jonathan Holdship}
\affiliation{Leiden Observatory, Leiden University, P.O. Box 9513, 2300 RA Leiden, The Netherlands}
\affiliation{Univsersity College London, Gower St, Bloomsbury, London WC1E 6BT, England}



\newcommand\NameOfCode{UCLCHEMCMC}
\begin{abstract}
    We present the publicly available, open source code \NameOfCode{}, designed to estimate physical parameters of an observed cloud of gas by combining Monte Carlo Markov Chain (MCMC) sampling with chemical and radiative transfer modeling. When given the observed values of different emission lines, \NameOfCode{} runs a Bayesian parameter inference, using a MCMC algorithm to sample the likelihood and produce an estimate of the posterior probability distribution of the parameters. \NameOfCode{} takes a full forward modeling approach, generating model observables from the physical parameters via chemical and radiative transfer modeling. While running \NameOfCode{}, the created chemical models and radiative transfer code results are stored in an SQL database, preventing redundant model calculations in future inferences. This means that the more \NameOfCode{} is used, the more efficient it becomes. Using UCLCHEM and RADEX, the increase of efficiency is nearly two orders of magnitude, going from $5185.33 \pm 1041.96  \, \mathrm{s}$ for ten walkers to take one thousand steps when the database is empty, to $68.89 \pm 45.39 \, \mathrm{s}$ when nearly all models requested are in the database. In order to demonstrate its usefulness we provide an example inference of \NameOfCode{} to estimate the physical parameters of mock data, and perform two inferences on the well studied prestellar core, L1544, one of which show that it is important to consider the substructures of an object when determining which emission lines to use.
\end{abstract}

\keywords{astrochemistry --- ISM: molecules}

%% cross-references to sections, equations, tables, and Figures.

\section{Introduction}\label{sec:Intro}
    Throughout the interstellar medium (ISM), chemical reactions impact the environments that we observe. In turn, the physics of a molecular cloud greatly affects the chemistry. For example, at high densities ($\gtrapprox \, 10^5\, \mathrm{cm}^{-3}$ ) and low temperatures ($\lessapprox \, 30 \, \mathrm{K}$), atoms and molecules freeze out onto dust grains where they can react through many pathways (for a review see \citet{Allodi}). On the other hand shocks from protostellar outflows impacting the surrounding medium can lead to desorption of many molecular species stored on dust grains \citep{Caselli_1997}. Hence, emission and absorption lines of  different species allow us to study the physics of the objects we observe. \par
    In order to interpret the observations that are made, radiative transfer codes can be used to calculate the expected intensities that should be observed with a given set of physical parameters. One example of such a code is RADEX \citep{radex} which focuses on non-LTE analysis. These types of codes require parameters that describe the condition of the gas as well as the column density of a species. These are connected by chemistry but treated as free parameters in many radiative transfer models. Modeling tools such as UCLCHEM \citep{uclchem} or GRAINOBLE \citep{GRAINOBLE} amongst many others, provide the fractional abundances of species which can be used to calculate the column density. The fractional abundances can be combined with estimates of the total gas column density of an object in order to calculate the column density of an individual species. \par
    \begin{figure*}[t]
        \includegraphics[width=\linewidth]{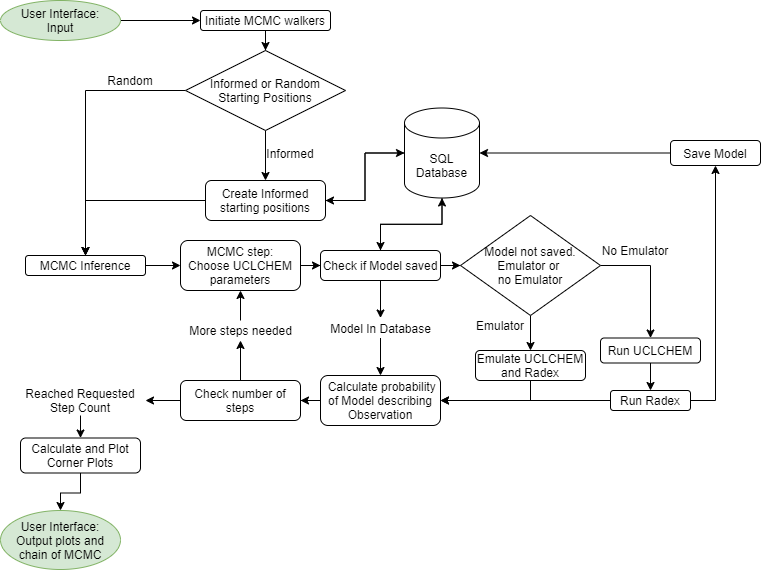}
        \caption{Flow Chart of the various processes that happen in \NameOfCode{}. Green ovals indicate parts that the User interfaces with. Diamonds indicate parts of \NameOfCode{} where the next step is dependent on options the user specifies. The SQL Database is represented with a cylinder and has been labeled this way for clarity. Arrows to and from the database represent a query of the SQL which then returns the models that match the query.}
        \label{Fig:FlowChart}
    \end{figure*}
    Chemical models calculate fractional abundances by considering the rate of change of many species as the interact through a network of reactions. These reaction networks usually include a gas phase database such as KIDA \citep{KIDA} or UMIST \citep{UMIST}, as well as gas-grain and grain surface processes such as freeze out, non thermal desorption and surface reactions. Additional processes such as thermal desorption or sputtering of ice mantles are often included depending on the chemical code and its intended purpose. The complexity involved in determining what should be included in these models in order to maximise the accuracy while minimising computational cost of creating a model, is an aspect that requires significant expertise. \par
    The best modeling approaches combine the chemical modeling codes with the radiative transfer codes. One benefit in doing this is that the column densities can be calculated with the chemical model which can then be combined with the set of physical parameters calculated by the chemical code to use as parameters for the radiative transfer model. There are additional parameters for the radiative transfer codes, such as line width, which are not directly calculated by a chemical code but can be treated as free parameters or derived from observations. The outputs from the radiative transfer code can then be compared to spectroscopic observations \citep{Harada, Punanova_2018, Viti_2017}. \par
    In order to assist in the inference of physical parameters of an observation, we present the open-source, MCMC inference tool \NameOfCode{}\footnote{\url{https://zenodo.org/badge/latestdoi/334982976}}. The intended use of \NameOfCode{} is to infer the probability distribution of key physical parameters given some observed data. The following section will describe the code in detail, starting with the forward modeling approach and what tools it uses in section~\ref{sub:ModT}, followed by the work flow of the code and how it stores models to allow future inferences to be more efficient in section~\ref{sub:sql} followed by a brief  description of the chosen interface in section~\ref{sub:input}. After that, we will examine an example case by providing \NameOfCode{} with mock observations, and then perform a stress test inference using observations of the prestellar core L1544 in section~\ref{sec:App}, before going on to discuss caveats for the use of this tool and summarising in section~\ref{sec:Sum}. \par
\section{\NameOfCode{}}\label{sec:Code}
    \begin{deluxetable*}{lll}[t]
        \label{tab:Options}
        \tablecaption{Inputs and options per page}
        \tablewidth{0pt}
        \tablehead{
        \colhead{Page} & \colhead{Input/Option} & \colhead{Description}
            }
        \startdata
            \makecell[l]{Parameter input \\ (Page 1)} & Final volume Density [cm$^3$] & Hydrogen volume density at which the model stops collapsing \\ 
            \hline
            & Kinetic temperature [K] &  Kinetic temperature of the gas \\ 
            \hline
            & Cosmic ray ionisation rate &  \makecell[l]{Multiplicative factor of the galactic rate of ionisation caused \\ by Cosmic rays ($1.3 \times 10^{-17} \, \mathrm{s}^{-1}$)} \\ 
            \hline
            & UV radiation field strength &  \makecell[l]{Strength of the external UV radiation field strength acting on \\ the cloud} \\
            \hline
            & R$_{\mathrm{out}}$ [pc] & Radius of the modelled cloud \\
            \hline
            & Line width [$\mathrm{km\, s}^{-1}$] & RADEX Line width of observation \\
            \hline
            \makecell[l]{Observation input \\ (Page 2)} & Species list &  List of the species that have been configured \\ 
            \hline
            & Transition list &  \makecell[l]{List of transition lines that have been configured for a given \\ species}\\
            \hline
            & Observation inputs & \makecell[l]{Space to fill in the observations, errors and \\ choice of units for the observation} \\
            \hline
            \makecell[l]{Options\\(Page 3)} & Grid type & \makecell[l]{Choice on whether to use coarse or fine grid \\ for the parameters being inferred} \\
            \hline
            & Informed starting position & \makecell[l]{Choice on whether to use informed starting \\ positions or random starting positions} \\
            \hline
            & Session name & \makecell[l]{Back-end session name to allow the session to \\ be reloaded later} \\
            \hline
            & Number of walker & \makecell[l]{Number of walkers the MCMC should use \\ (It is recommended by the package emcee to use twice as \\ many walkers as parameters being inferred)} \\
            \hline
            & Number of steps & \makecell[l]{choice of the number of steps an inference \\ should take before stopping and loading evaluation corner \\ plots} \\
            \hline
            \makecell[l]{Inference\\(Page 4)}& Start inference & \makecell[l]{ to start the MCMC inference or continue it \\ if a previous session is loaded} \\
            \hline
            & MCMC corner plot & \makecell[l]{MCMC corner plots of previous steps, if \\ previous steps exist, or the starting positions if a new inference \\ is being started} \\
            \hline
        \enddata
        \tablecomments{Options of inputs and outputs per page for \NameOfCode{}}
    \end{deluxetable*}
    \begin{figure*}[t]
        \centering
        \includegraphics[width=\linewidth]{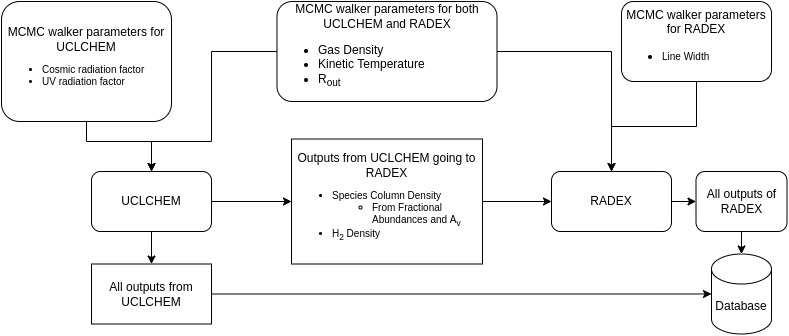}
        \caption{Small flow chart showing which parameters go into UCLCHEM and which parameters are taken from UCLCHEM and given to RADEX.}
        \label{fig:UCLCHEM-RAD}
    \end{figure*}
    \NameOfCode{} infers physical parameter values from molecular observations using chemical and radiative transfer models. First, we use a chemical model, in order to obtain abundances for a user-defined list of species \NameOfCode{} has been configured for. These abundances, can then be used with a radiative transfer model, to calculate the intensities for the emission lines of those species. For a single model, this process can take several minutes to be calculated on a standard computer. \par 
    In order to infer the physical parameter values, we use the affine-invariant Markov Chain Monte Carlo (MCMC) Ensemble algorithm of \citet{Good} as implemented in the python package emcee \citep{emcee}. This kind of sampling initiates walkers with a set of physical parameters which are used to calculate the chemical and radiative transfer models as just described. During each step, the walkers calculate the likelihood value for that set of parameters using the likelihood function given by the user (see section~\ref{sub:mcmc} for the details on the likelihood). After calculating the likelihood of the current values, the walkers choose a new set of values in parameter space, for which the likelihood is also evaluated. At this point, the walker must decide if it will remain stationary for this step, discarding the new set of values and keeping the one it already has, or if it will discard the old values and keep the new ones. This decision is dependent on the type of "move" function that is chosen (for details, please refer to \citet{emcee}). The default function for \NameOfCode{} is a combination of a differential evolution proposal \citep{BenNelson} and a snooker proposal using differential evolution \citep{Braak}. Mixing like this is recommended by \citet{emcee} when dealing with multi-modal problems as the differential evolution proposal can allow for large enough step sizes to cross between peaks in probability. However, the standard version strongly recommends that at least $N=2d$ walkers are used, where $d$ is the number of parameters over which to infer, and performs better with higher $N$. The snooker proposal using differential evolution allows for improved performance, compared to the basic version, when a smaller number of walkers is used. The process of sampling parameter space in this way requires thousands of models to be calculated before a meaningful posterior is produced. The calculations of each model using UCLCHEM and RADEX can take a minute or more, which can result in several hours of computing time prior to the parameter space being sampled enough to produce a usable posterior. \par
    Our aim is to improve the efficiency without decreasing the accuracy. To do this \NameOfCode{} manages a database of previously calculated models from which it can retrieve values when required. The curated database contains the input and output from both the chemical and radiative transfer models, and is used to perform an inference of the physical parameter space of an observed object without repeating any calculation. Anytime a new step is taken, it can check if this combination of parameters has been used before, and if it has, use the old output rather than perform a new calculation. A full flowchart of the processes done can be found in Figure~\ref{Fig:FlowChart}. In this section we will start by briefly describing the forward modeling method we use and the software \NameOfCode{} requires in order to create the models that it stores, followed by the details of the MCMC inference and how \NameOfCode{} manages the database, before detailing the interface it has. \par

    \subsection{Forward modeling}\label{sub:ModT}
        To create the simulations that are stored in the database, we use UCLCHEM combined with RADEX as the chemical model and radiative transfer code respectively. Physical parameters describing the gas conditions are passed to UCLCHEM to generate abundances and then a subset of these values are passed to RADEX in order to get a list of transition lines. Beyond the outputs from UCLCHEM, additional free parameters are required for RADEX such as the line width. A detailed flow chart on how the modeling tools interact with each other can be found in Figure~\ref{fig:UCLCHEM-RAD}, for clarity. The inputs for UCLCHEM and RADEX, listed in the flow chart, can be changed according to the needs of the inference to be run, but requires changes to be made to configuration files. \par
        By default, \NameOfCode{} is configured to use RADEX for the radiative transfer calculations; we use the fractional abundance, calculated by UCLCHEM, to approximate the column density by using the visual extinction calculated from the given $\mathrm{R}_{\mathrm{out}}$ and gas volume density. This value is used along side the other inputs UCLCHEM was given, that RADEX needs as well, such as gas volume density and kinetic temperature, in order to run the radiative transfer model to produce observables which can be compared to the data. The observable values RADEX produces are: radiation peak temperature ($\mathrm{T}_{\mathrm{R}}$) in K which is comparable to the measured line intensity; integrated surface brightness in $\mathrm{K\,km}\,\mathrm{s}^{-1}$; the isotopic flux emitted in all directions in $\mathrm{ergs}\,(\mathrm{s\,cm}^2)^{-1}$ \citep{radex}. Any of these can be sent to the likelihood function depending on the user's data. \par
        Both UCLCHEM and RADEX can be replaced, as \NameOfCode{} is only designed to perform an MCMC inference and manage an SQL database. As long as inputs and outputs are carefully tailored to a given project, the code could be simply modified to be used with any chemical modeling or radiative transfer codes. To begin, we use a limited list of chemical species whose collisional data is available in the Leiden Atomic and Molecular Database (LAMDA \citet{lambda}), as RADEX requires such data. \par
    \subsection{MCMC Inference}\label{sub:mcmc}
        For the MCMC inference, \NameOfCode{} uses the python package emcee \citep{emcee} in order to calculate the posterior probability density function (PDF) of the physical parameters for the desired observation. We assume the errors on the data are Gaussian and that our model provides the true intensities for any given parameters. The initially configured likelihood of observing our data given some parameters that was used for the example cases in section \ref{sec:App} is therefore,
        \begin{equation}\label{eq:like}
            \mathcal{L}(\mathrm{d}|\,\theta)=
            \exp \left[-\frac{1}{2}\sum_i\frac{(\mathrm{d}_i-\theta_i)^2}{\sigma_{\mathrm{d}_i}^2}\right],
        \end{equation}
        where $\mathrm{d}_i$ represents the observed input, $\theta_i$ the output from RADEX, using physical parameters from UCLCHEM, and $\sigma_{\mathrm{d}_i}$ is the error in the observed input, each for line $i$. This is then combined with a prior on the physical parameters, $\mathrm{P}(\theta)$ and the Bayesian evidence, $\mathrm{P}(\mathrm{d})$, in Bayes theorem to get the full PDF in the form of
        \begin{equation}\label{eq:PDF}
            \mathrm{P}(\theta | \, \mathrm{d})  = \frac{\mathrm{P}(\theta) \, \mathcal{L}(\mathrm{d}|\,\theta)}{\mathrm{P}(\mathrm{d})}.
        \end{equation}
        By default, the prior is a uniform top hat function in grid space, on the ranges designated by the end user but can be altered by end users. As the prior is applied in grid space, it will be a log uniform prior if the physical parameter has a log spaced grid applied to it. The evidence is treated as a normalisation factor, and the values that are used for our example application on prestellar core will be discussed in section~\ref{sec:App}. This approach to parameter inference has been used before \citep{Holdship_2019} and UCLCHEMCMC makes such inference problems simple. \par

    \subsection{Database}\label{sub:sql}
        The core of \NameOfCode{} is the managing of a database of models and running an MCMC inference which is supplemented by that database. After giving the inputs, which will be detailed in section~\ref{sub:input}, \NameOfCode{} initiates a string of operations in order to start calculating the posterior PDF of the physical parameter values of an object using an MCMC sampler. A detailed flow chart of the processes can be seen in Figure~\ref{Fig:FlowChart}. The code starts by initiating the walkers that will be used for the MCMC inference. If informed starting positions were requested, the SQL database will be searched for models which are similar to the given observations based on a simple top-hat function with a configurable distance on either side of the observed intensities. By searching and retrieving the parameters of models that have intensities similar to the given observation, we can construct a function by calculating the mean and standard deviation for each parameter to produce a normal distribution from which we can sample starting positions. Each parameter will have its own distribution from which the starting positions for the MCMC walkers are sampled. If random starting positions are chosen, then a uniform distribution of the parameter space is sampled in order to create the starting positions for each walker. Upon creating the starting positions, we then invoke the MCMC inference which will need to be able to access the SQL database as described in section~\ref{sub:sql}.\par
        For the sake of storing the inputs and outputs from UCLCHEM and RADEX, we use an SQL database using the SQLite implementation. This is chosen as it is a light weight, widely used, and easy to implement solution for storing large volumes of data in such a way that it can easily be queried. The main advantage of having access to an SQL search method is that it can quickly check if the combination of parameters to be calculated has been previously stored. If it has, then the program goes directly to evaluating the likelihood using equation~\ref{eq:like}, which takes an almost negligible amount of time to perform. If the combination of parameters is not in the database, then a model can be created and stored for that set of parameters. This means that the calculation speed of the MCMC inference is dependent not only on the number of walkers and desired steps, but also on how many models are stored in the database. \par 
        Based on that, \NameOfCode{} will become faster as more inferences are performed. The calculation of a single model can take around one minute when using UCLCHEM and RADEX. While this can be parallelised, it is still a limiting factor when thousands of models have to be calculated to get a reasonable estimation of a PDF. On the contrary, the action of submitting a query to an SQL database and evaluating the probability of the stored models matching the observations takes less than a second. Improving efficiency this way has the advantage over techniques such as emulation \citep{uclchem_em} as no approximation is made. To quantify the improvement, we measure the time it took ten walkers, to perform one hundred steps, at three different times: (i) When no database is being used; (ii) When around half of the models the inference wants to use are retrieved from the database; (iii) When nearly all models the inference is using are retrieved from the database. We use this type of measurement for the performance, as the minimum time, the time when every model can be retrieved from the database, should be identical irrespective of the chemical and radiative transfer model that is used. For the three cases, the mean time and standard deviation are: (i) $5185.33\, \mathrm{s} \pm 1041.96\, \mathrm{s}$; (ii) $4834.67 \, \mathrm{s} \pm 843.24 \, \mathrm{s}$; (iii) $68.89 \, \mathrm{s} \pm 45.39\, \mathrm{s}$. We emphasise that the times found for case (i) and (ii) are strongly dependent on the chemical and radiative transfer models that are used, while case (iii) should only be weakly dependent on which models are used, with the dependency disappearing if all models the inference wants to use are within the database. \par
        The database can be accessed both by the code, and by a user who wishes to use the models stored within it for other purposes. At the time of the first release, we store all inputs that are given to the chemical model when it is run, as well as all outputs that are produced by it. This can include the output of intermediate time steps which UCLCHEMCMC can be set up to store if requested. \NameOfCode{} then takes the given line width, either as a free parameter of the MCMC or as a constant value given by the user, as well as the kinetic temperature, volume density of H$_2$, and the fractional abundances of the atomic or molecular species from the chemical code in order to run the radiative transfer code. The outputs from this code are then stored in the SQL database. From  here, the emission lines given by the radiative transfer code can be compared to the observations to evaluate the likelihood as discussed previously. \par 
        
    \subsection{Interface}\label{sub:input}
        The User Interface (UI) for \NameOfCode{} is browser based, in order to give a simple usable interface that should be compatible with most operating systems. A further advantage of this is that an online, publicly available version can be more easily created in the future such that end users will not need to change the workflow. \par
        The inputs that are requested from a user of \NameOfCode{} are separated onto three pages within the UI. The first page requests the ranges of the physical parameters over which the inference should be performed. At the time of the first release, the configured parameters are: (i) the volume density of the gas in cm$^{-3}$ at which point the model should stop collapsing; (ii) the kinetic temperature of the gas in K; (iii) the cosmic ray (CR) ionisation rate in units of the galactic CR ionisation rate ($\zeta_0$); (iv) UV radiation field strength in units of Habing; (v) the radius of the assumed spherical cloud being modelled in parsec (R$_{\mathrm{out}}$); (vi) the line width to be used with RADEX in units of $\mathrm{km \, s}^{-1}$. Upon supplying the desired ranges and which parameters should be kept constant, the next set of inputs is the observations. Here, a list of species can be selected to be added to the current inference. Once a species has been selected, the compatible lines will be shown and can be selected, after which it is possible to add the values of the observations, errors and the observed quantity. As of the first release, \NameOfCode{} is configured to allow for the units that RADEX has as outputs, detailed in section~\ref{sub:ModT}.\par
        The penultimate page contains the options for the MCMC inference. The options are: the MCMC details, walker starting positions, and grid type. There are three options for the MCMC algorithm that an end user can easily change and they are: (i) the number of walkers that the inference should have; (ii) the number of steps the inference should perform before saving; (iii) the name of the session. Naming the session allows for an inference to be started again at a later time without having to re-enter all the previous parameters and observations. This was added, in case the code crashes, or if after evaluating the results it was determined that the MCMC walkers could benefit from more steps to ensure the walkers converged. The starting positions of the MCMC walkers can either be randomly determined or set to inform starting positions depending on the end users preference. The details of how informed starting positions are calculated are given in section~\ref{sub:sql}. The grid type option allows an end user to choose which physical parameter space grid they want to use for the inference they are going to run, and are intended to be created and managed by the end user. By default, there is a coarse and fine grid provided to give an example of how they are meant to be created. The discretisation of the parameter space to grids was implemented as chemical models with physical parameters that differ only to a small degree would produce nearly indistinguishable outputs but would be considered separate models by the code that retrieves and stores models in the SQL database. This would lead to models with minor differences in parameters space being calculated despite producing indistinguishable outputs. \par

\section{Application}\label{sec:App}
    \begin{deluxetable}{ccc}[t]
        \label{tab:ParamRanges}
        \tablecaption{Parameter Ranges}
        \tablewidth{\linewidth}
        \tablehead{
        \colhead{Parameter [Units]} & \colhead{Lower Bound} & \colhead{Upper Bound}}
        \startdata
            Volume Density [cm$^{-3}$] & $5.0 \times 10^{4}$ & $1.0 \times 10^{7}$ \\
            Kinetic Temperature [K] & $5$ & $20$ \\
            UV radiation field [Habing] & $0.1$ & $10$ \\
            R$_{\mathrm{out}}$ [pc] & $0.0001$ & $0.1$ \\
            CR ionisation rate [$1.3 \times 10^{-17} \mathrm{s}^{-1}$] & $0.1$ & $10$
        \enddata
        \tablecomments{Physical Parameter range the inference is allowed to explore for both the mock and observational inference.}
    \end{deluxetable}
    \begin{deluxetable*}{ccccccc}[t]
        \label{tab:Synthetic}
        \tablecaption{Mock Data used for Evaluation}
        \tablewidth{0pt}
        \tablehead{
        \colhead{Species} & \colhead{Transition} & \colhead{Frequency [GHz]} & Line width [$km\,s^{-1}$] & \colhead{RADEX Value} & \colhead{Mock Data (T$_\mathrm{MB}$)} & \colhead{Units}
        }
        \startdata
            CS & 2,0 - 1,0 & 97.98095 & 1.0 & 2790.9 & $2412.4 \pm 558.2$ & mK \\
            SO & 2,2 - 1,1 & 86.09395 & 1.0 & 1918.6 & $2553.0 \pm 383.7$ & mK \\
            & 2,3 - 1,2 & 99.29987 & 1.0 & 450.3 & $367.6 \pm 90.0$ & mK \\
            & 3,1 - 2,1 & 109.2522 & 1.0 & 185.5 & $222.3 \pm 37.1$ & mK \\
            o-H$_2$CS & $3_{1,3}-2_{1,2}$ & 101.4778 & 1.0  & 130.6 & $151.6 \pm 26.1$ & mK \\
            & $3_{1,2}-2_{1,1}$ & 104.6170 & 1.0  & 85.5 & $83.8 \pm 17$ & mK \\
        \enddata
        \tablecomments{Values of the mock data created for evaluation of \NameOfCode{} using UCLCHEM and RADEX. The RADEX Value column contains the value given by RADEX, while the Mock Data column contains the same values with added Gaussian noise, and corresponding error values.}
    \end{deluxetable*}
    \begin{deluxetable*}{cccccc}[t]
        \label{tab:Lines}
        \tablecaption{Observations used for evaluation}
        \tablewidth{0pt}
        \tablehead{
            \colhead{Species} & \colhead{Transition} & \colhead{Frequency [GHz]} & Line width [$km\,s^{-1}$] & \colhead{Observation [T$_\mathrm{mb}$]} & \colhead{Units}
        }
        \startdata
            CS & 2,0 - 1,0 & 97.98095 & 0.64$\pm$0.07 & $1226.5 \pm 0.1$ & mK\\
            SO & 2,2 - 1,1 & 86.09395 & 0.42$\pm$0.01 & $223.7 \pm 5.9$ & mK\\
            & 2,3 - 1,2 & 99.29987 & 0.45$\pm$0.01 & $1422.5 \pm 40.6$ & mK\\
            & 3,1 - 2,1 & 109.2522 & 0.39$\pm$0.01 & $176.1 \pm 5.2$ & mK\\
            HCS+ & 2-1 & 85.34789 & 0.43$\pm$0.01 & $246.8 \pm 6.1$ & mK\\
            OCS & 6-5 & 72.97678 & 0.36$\pm$0.01 & $106.3 \pm 6.8$ & mK\\
            & 7-6 & 85.13910 & 0.38$\pm$0.01 & $87.4 \pm 7.2$ & mK\\
            & 8-7 & 97.30121 & 0.37$\pm$0.01 & $70.5 \pm 5.4$ & mK\\
            & 9-8 & 109.4631 & 0.34$\pm$0.04 & $49.4 \pm 6.4$ & mK\\
            o-H$_2$CS & $3_{1,3}-2_{1,2}$ & 101.4778 & 0.44$\pm$0.01 & $558.4 \pm 11.7$ & mK\\
            & $3_{1,2}-2_{1,1}$ & 104.6170 & 0.44$\pm$0.01 & $514.3 \pm 12.0$ & mK\\
            p-H$_2$CS & $3_{0,3}-2_{0,2}$ & 103.0405 & 0.45$\pm$0.02 & $536.9 \pm 16.5$ & mK\\
        \enddata
        \tablecomments{Observations collected from \citep{Vastel_2014} and \citep{Vastel_2018}, "o-" and "p-" represent ortho- and para- version of species respectively.}
    \end{deluxetable*}
    In order to give an example of \NameOfCode{}, we run three inferences. One inference is on mock data which were created by using UCLCHEM and RADEX, as these two codes are used in \NameOfCode{} to perform the inference. The second and third inference are for the prestellar core L1544 \citep{Caselli_2002}, once considering the emission from only one molecular species and once with all sulfur bearing species. The data we used for L1544 can be found in table~\ref{tab:Lines} and were used to infer the kinetic temperature, volume density, CR ionisation rate, and R$_{out}$. This object is a very well studied prestellar core located at R.A. = $05^{\mathrm{h}}01^{\mathrm{m}}11^{\mathrm{s}}.0$, Dec=$25^{\circ}07'00''$ \citep{Caselli_2002} (\citet{Vastel_2014}, \citet{Punanova_2018} and \citet{Vastel_2018}). \par
    We use the same input parameter space for all inferences. The exception is the UV radiation field which we hold constant for the inferences on L1544 as we expect the visual extinction to be sufficiently high for changes in the UV to be negligible. The ranges for the physical parameters can be found in table~\ref{tab:ParamRanges}.
    
    \subsection{Mock Data Inference}
        \begin{figure*}
            \centering
            \includegraphics[width=\linewidth]{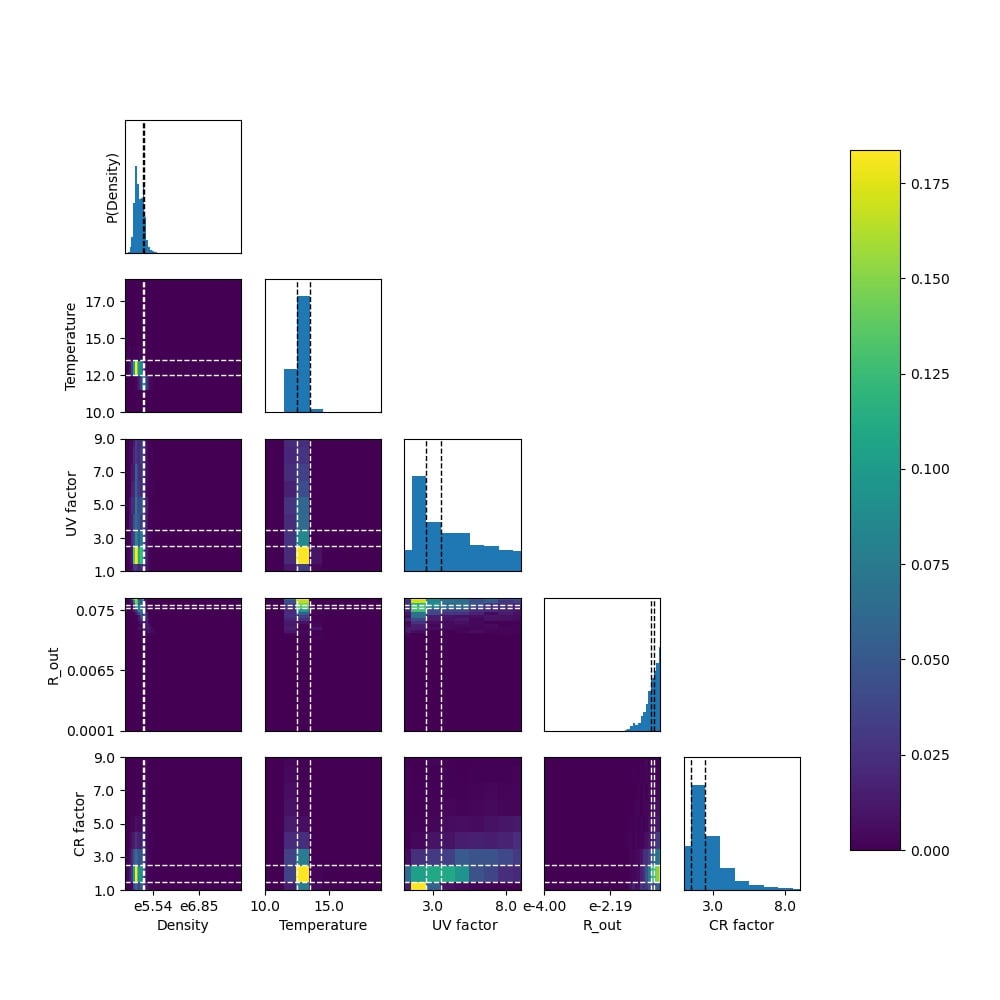}
            \caption{Posterior distribution function of the evaluation run performed on mock data. The histograms represent the PDF of volume density, kinetic temperature, UV field factor, R$_{\mathrm{out}}$ and the cosmic ray ionisation rate factor, the colour bar shows the value ranges of the joint distribution functions. The white dashed lines in the joint distributions, and the black dashed lines in the PDFs represent the true value used to create the mock data. }
            \label{fig:Synthetic}
        \end{figure*}
        First, we verify that \NameOfCode{} performs as intended by creating mock data using the same modeling codes that \NameOfCode{} uses to perform an inference. We add Gaussian noise to the data with a standard deviation of five percent for each emission line as this would make the mock data errors on par, but slightly higher than, the average of the uncertainties on the L1544 observational data. We do this because running an inference where the true values are known and the data are model generated allows us to test whether \NameOfCode{} performs as intended when the models are appropriate for to the data. As this is just an example case, and many different combinations of chemicals and transition lines could be picked, we choose a subset of emission lines from the observations we use for the inferences on L1544. In order to create the mock data, we randomly chose physical parameters that resulted in all emission lines having an observable flux. The parameters we chose are as follows: Final volume density $1.0 \times 10^5\,[\mathrm{cm}^{-3}]$, a kinetic gas temperature of 13 [K],  radiation field of 3 Habing, a cloud radius of $0.08\, \mathrm{pc}$, and a CR ionisation rate value of $2.6 \times 10^{-17} \, \mathrm{s}^{-1}$. The emission lines and corresponding mock data values are found in table~\ref{tab:Synthetic}, which contains the exact values of each line given by UCLCHEM and RADEX prior to adding noise, as well as the data with Gaussian noise added to it and the corresponding uncertainties.\par
        \begin{figure}
            \centering
            \includegraphics[width=\linewidth]{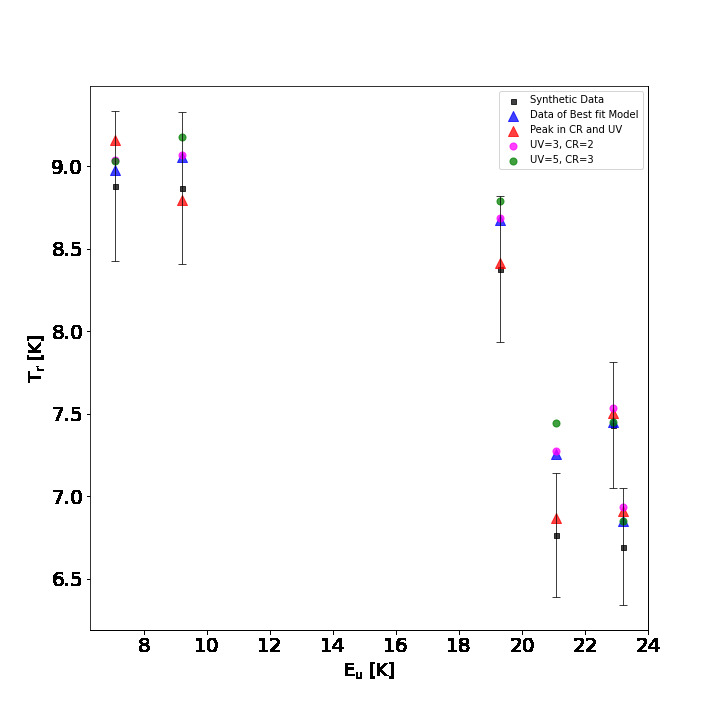}
            \caption{The radiation temperature,T$_\mathrm{R}$, calculated by RADEX against the Energy of the upper state, for the mock data and errors given to \NameOfCode{} in black, and the data created when using the most likely parameter values from the 1D distributions from the inference of the mock data in blue. Red represents the peak in the joint distribution of the CR ionisation rate and UV radiation field while keeping the remaining parameters as they are for the previous model, while green and fuchsia represent two additional points with values for the CR ionisation rate and UV radiation field values in the elongated distribution of likely values to show why the inference still gave some importance to these values.}
            \label{fig:SyntheticRotation}
        \end{figure}
        We run the inference, monitoring the chain of steps that each walker has taken. The likelihood of accepting a new set of parameters decreases as a function of the difference between the likelihood of the current model and the new model. This means that over time, the parameter space that is being traversed by all walkers, will decrease as the walkers find areas of parameter space where the set of parameters produce models with a higher likelihood. Once the walkers stop reducing this parameter space, we stop the inference. Using these chains from the inference, we then create the posterior, shown in Figure~\ref{fig:Synthetic}. The distributions contain the true values, which indicates that \NameOfCode{} works as expected. In order to validate this, we plot all mock observation lines against the upper state energy, seen in Figure~\ref{fig:SyntheticRotation}, and do the same for the model values that are produced from \NameOfCode{}'s parameters with the highest likelihoods in the 1D distributions. When we do this, we see that all but one line lies within the uncertainties of the mock data. \par
        We note that in the posteriors, there is a considerable degeneracy between the UV field and the CR ionisation rates. Additionally, there is a clear peak in the joint distribution of CR ionisation rate and UV radiation field. This peak is at a CR ionisation rate equal to the galactic value (CR=1) and at a UV radiation field strength of two Habing, however this peak does not have the same value of the CR ionisation rate as the parameter set used to generate Figure~\ref{fig:SyntheticRotation} as that takes the most likely value from the 1D marginalised distributions rather than the overall most likely parameter set. To see how the emission lines change along this extended distribution, and how the observations look at this peak in the joint distribution, we include the emission lines of this peak, and two additional combinations of CR ionisation rate and UV radiation field strength in Figure~\ref{fig:SyntheticRotation}, while holding all other parameters constant. In looking at how the antenna temperature of the emission lines compare between the models and the mock data, it becomes quite clear that the values of the observations in this distribution all show significant agreement with the mock data but that the peak in the CR ionisation rate and UV radiation field strength distribution produces observations that fit better than the model created using the most likely parameter values in the 1D distribution. 
    \subsection{Inferring the physical parameters of the pre-stellar core L1544}\label{sub:L1544}
        \begin{figure*}
            \centering
            \includegraphics[width=\linewidth]{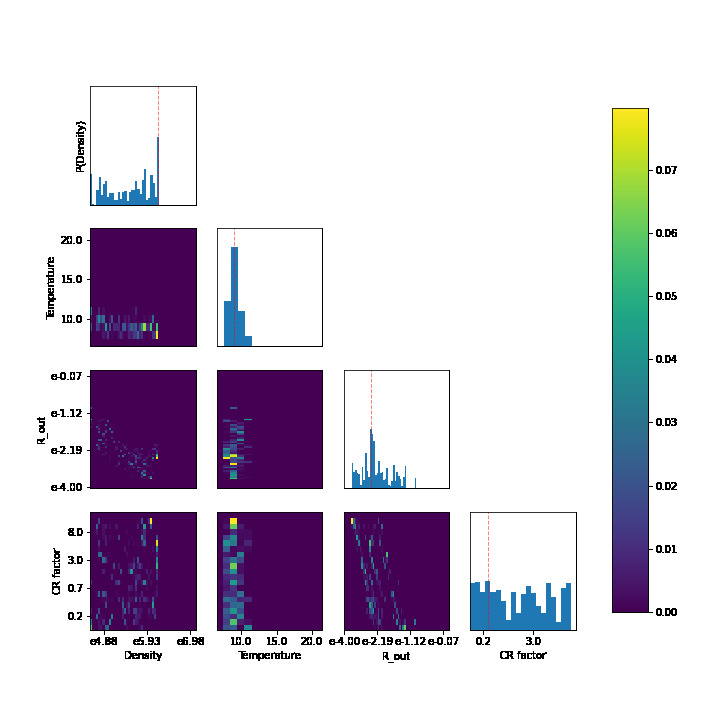}
            \caption{Posterior distribution function of the evaluation run performed on the emission lines from OCS only. The histograms represent the PDF of volume density, kinetic temperature, R$_{\mathrm{out}}$ and the cosmic ray ionisation rate factor, the colour bar shows the value ranges of the joint distribution functions, while the red dashed line in the PDF is the value with the highest probability.}
            \label{fig:OCSInference}
        \end{figure*}
        \begin{figure}
            \centering
            \includegraphics[width=\linewidth]{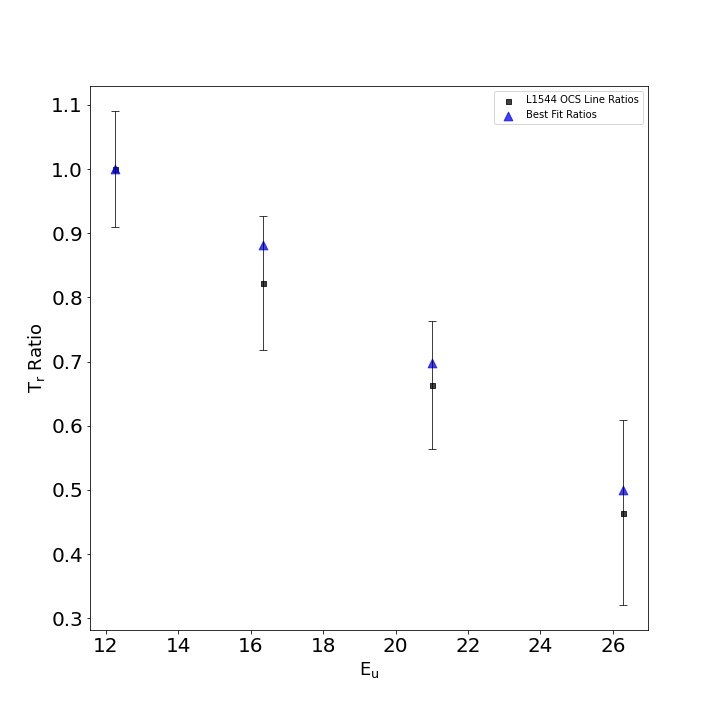}
            \caption{T$_\mathrm{R}$ over T$_\mathrm{R}$ of OCS 6-5 against the upper state energy, for emission lines of OCS found in table~\ref{tab:Lines}, compared to the data of the best fit model after running an inference using only the OCS lines. All of the lines fit the observed line ratios quite well. Black represents the real data with error bars, while blue is the best fit model.}
            \label{fig:OCSRotation}
        \end{figure}
%        \z\z\xyz
        \begin{figure}
            \centering
            \includegraphics[width=\linewidth]{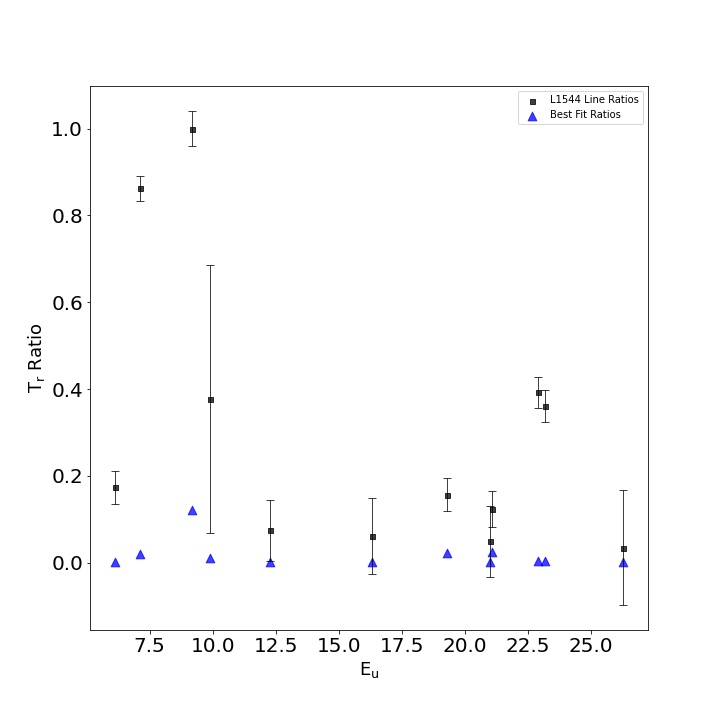}
            \caption{T$_\mathrm{R}$ against the Energy of the upper state, for all emission lines in table~\ref{tab:Lines}, compared to the data of the best fit model after running the stress test inference. While a couple lines almost fit their observed counterparts, it is clear that \NameOfCode{} is unable to match all lines at once, which made it settle for a set of parameters, that allow each line to at least get somewhat close to the observations. Black represents the real data with error bars, while blue is the best fit model.}
            \label{fig:FullRotation}
        \end{figure}
%        \xyz\z\xyz\z
        \begin{figure*}
            \centering
            \includegraphics[width=\linewidth]{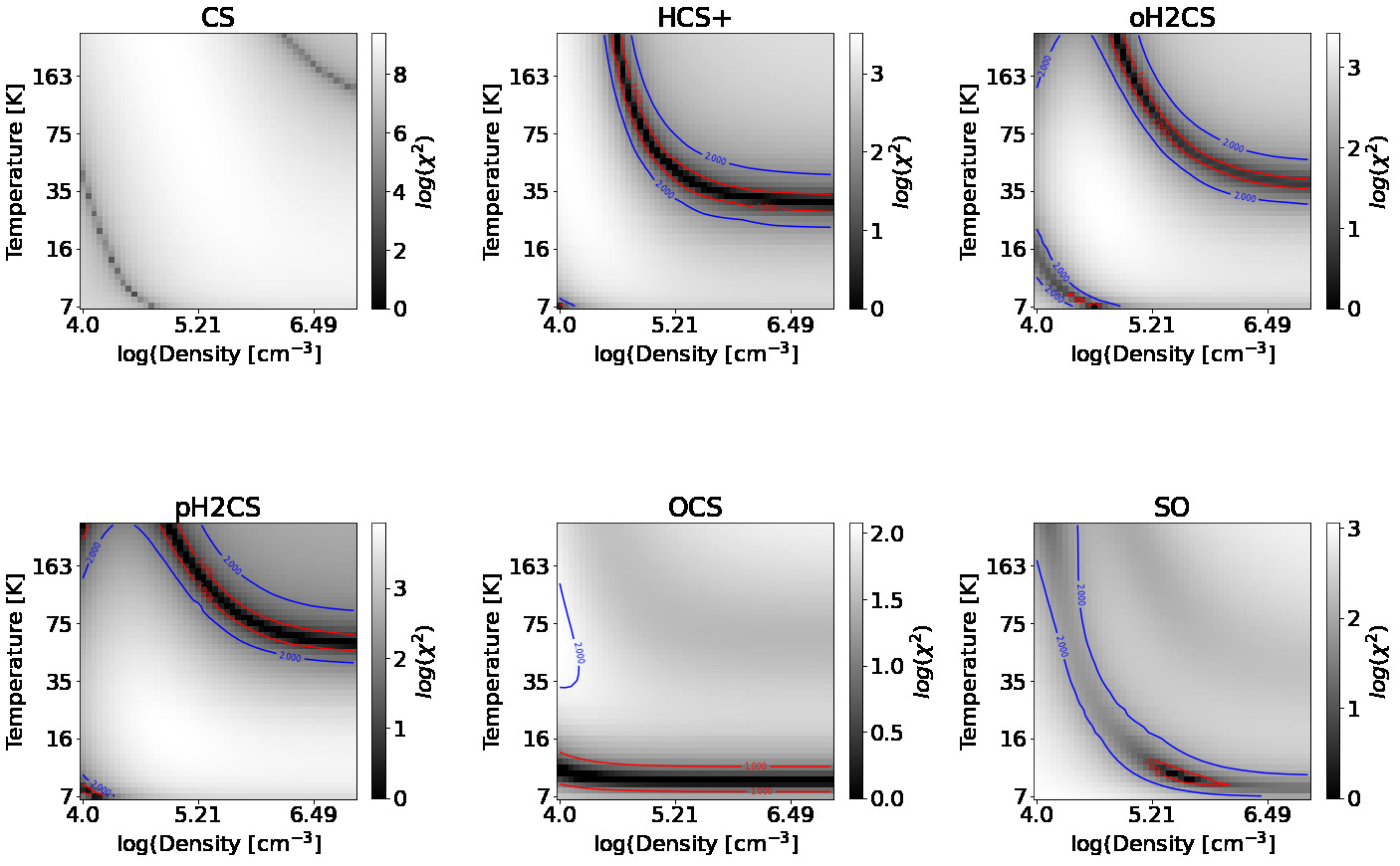}
            \caption{log($\chi^2$) grid for kinetic temperature and volume density using column density from \citet{Vastel_2018} for the six species that are used for the MCMC Inference. The lower value of the log($\chi^2$) is fixed at 0 to allow for better comparison between each species while allowing a flexible upper end, as the large ranges of log($\chi^2$) values make it difficult to make an informative Figure with a single range of values.}
            \label{fig:RadexOnlyFit}
        \end{figure*}
        With verification of how well \NameOfCode{} can perform when using mock data, we now use real observations in order to run another inference. The observations we use are from \citet{Vastel_2018}, which which present observations for two distinct regions in the L1544 object. One region is a methanol shell that is around 8000 AU \citep{Vastel_2014} from the core and in this shell UV photons can desorb methanol from the dust. The other region is a dust peak situated at the center of the object. We perform two inferences on the dust peak as there is a collection of sulphur bearing species where we start by using only the emission lines of OCS. In this first inference, we start with very broad priors and do not include additional information on the priors to show how \NameOfCode{} can perform. In an inference that is not designed to showcase the performance, any prior knowledge would be used to inform the ranges of the priors. We then follow that with an inference using all sulphur bearing species, to serve as a test to inform us of how well of an inference can be performed when \NameOfCode{} is given unfavourable conditions. More chemical species and emission lines are potentially unfavourable conditions, as \NameOfCode{}, may  struggle to find a single set of parameters, that fit all lines at once. For reference of values we expect the inference to estimate, we turn to \citet{Vastel_2018} who model the kinetic temperature and the volume density of the dust peak to be around $~7 \mathrm{K}$ and $~ 2 \times 10^6 \mathrm{cm}^{-3}$ respectively. \par
        As this is an example case of how to use \NameOfCode{}, we leave a large physical parameter range for the volume density, R$_{\mathrm{out}}$, and CR ionisation rate value. We limit the kinetic temperature to be between 5 and 30 K, as at this temperature, a single degree can make a difference to the diffusion and desorption rates of various species, impacting the fractional abundances of different species. The only two exceptions we make on limiting parameters is leaving the UV radiation field strength at the default value for UCLCHEM, as it is not a parameter of interest for the dust peak of a pre-stellar core, and setting the line width to the error weighted mean of $0.37 km \, s^{-1}$ for the OCS only inference. \par 
        We follow this, by using all of the chemical lines found in table~\ref{tab:Lines} for a second inference of the dust peak. We intentionally use all of these lines as a stress test. For the second inference, we make the assumption that these species trace the same substructure, which we emphasise in the inference by setting all line widths to $1.0 km \, s^{-1}$ in \NameOfCode{}. This could lead to an inference that is unable to fit the observations as \NameOfCode{} could struggle to find one set of parameters that lead to emission lines that match the observations. \par
        The posterior of the limited inference, can be found in Figure~\ref{fig:OCSInference}. The distribution has a very broad range in volume density and the CR ionisation rate value, while having a strong peak in kinetic temperature and peaked area in R$_{\mathrm{out}}$. This suggests that there is a wide range of possible volume density and CR ionisation rate values that can describe the observations well.\par
        In order to choose a good fit, we use previously modeled values of the total column density along with the fractional abundance of hydrogen to constrain the gas volume density. \citet{Caselli_2002} modeled a total column density of $4.4 \times 10^{22}\,[\mathrm{cm}^{-2}]$ which we combine with the peak value of the R$_{\mathrm{out}}$ posterior to obtain a likely volume density of $10^6\,[\mathrm{cm}^{-3}]$. \par
        To validate that this is a good fit, we plot the emission line ratios, with respect to the most intense line OCS 6-5, against the upper state energy, to get Figure~\ref{fig:OCSRotation} to create a diagram analogous to a rotation diagram. This diagram shows that the modelled line ratios, are within the estimated error bars of the observed line ratios, which supports the accuracy of the inference performed. A useful next step would be to remove volume density as a free parameter, and use the measured column density to calculate it from R$_{\mathrm{out}}$ during another round of inference, with a finer grid in parameter space. Since this is just an example case, we will instead move on to the stress test of \NameOfCode{}. \par 
        Prior to running this test, we calculate a $\chi^2$ fit of volume density and kinetic temperature using only the radiative transfer code RADEX, to serve as a baseline comparison to the performance of \NameOfCode{} on the observed data, as this is a more common approach. To perform this fit, we take the column densities, determined by \citet{Vastel_2018} through radiative transfer modeling, for each species in table~\ref{tab:Lines}, and run RADEX on a grid of kinetic temperatures and volume densities. Results of this fit can be seen in Figure~\ref{fig:RadexOnlyFit}. The $\chi^2$ fit is unable to find one set of parameters that agree with each other for all lines. It also accepts a very large area of parameter space as potential fits to each individual species, severely limiting how helpful this fit is to any modeling effort. Beyond that, this method requires that we either provide a column density estimate, or that we include a grid of column densities over which to calculate, which would significantly increase the calculation time as it would be adding an additional dimension to the parameter space. We note however, that the speed at which these calculations was performed is at least three orders of magnitude lower than a traditional MCMC inference that does not use an SQL database. \par 
        We perform the stress test inference with the observational data by using all species and emission lines found in table~\ref{tab:Lines}. As is the case with the $\chi^2$ fit, the stress test inference is unable to match all lines at once. The area onto which the MCMC inference converges is a delta-like distribution on a single set of parameters that we then use to model the emission lines. We again plot the emission line ratios against the upper state energy, this time with respect to the SO (2,3)-(1,2) line as it is the strongest line, resulting in Figure~\ref{fig:FullRotation}. In this Figure it is clear that while one or two of the line ratios fit the observed ratios, the vast majority of lines from the best fit model do not match the observations at all. This failure is expected as \NameOfCode{} assumes a simple homogeneous model should fit the observations. As more species and transitions are added, the assumption of a simple homogeneous model will be broken. We include this example of a failed fit, to assist users of \NameOfCode{} in understanding some of the limitations and more importantly as a cautionary note when trying to fit multiple molecular transitions with one single gas component. Beyond that, it is important to analyse the best fit models and not just assume the posterior must be a good fit.\par
\section{Summary}\label{sec:Sum}
    The publicly available MCMC inference and SQL database managing tool, \NameOfCode{}, is capable of inferring physical parameters of astrochemical observations. This paper presents the details necessary to understand the use, strengths and shortcomings of this tool. The management of the database, using SQLite, increases the efficiency of parameter inference as the tool is used. Using the MCMC inference package, emcee, as well as having decoupled the chemical code and radiative transfer code from the inference, also makes \NameOfCode{} capable of handling any other chemical modeling or radiative transfer modeling tool. \par
    We showed the outputs of \NameOfCode{} when inferring the physical parameters of mock data created using UCLCHEM and RADEX, detailing just how well this recovered the physical parameters of the mock data. The use of the SQL database in the inference has showed that, once most models the inference looks for are in the database,  \NameOfCode{} goes from taking $5185.33 \pm 1041.96  \, \mathrm{s}$ for ten walkers to take one thousand steps to needing $68.89 \pm 45.39 \, \mathrm{s}$ which is a significant decrease in computational time. When inferring the physical parameters of actual observations, we detailed some of the issues that must be taken into consideration when running this tool. Users should be aware that if they keep the physical parameter ranges too small, then the inference may not be able to find matching parameters, resulting in non physical answers. We intend to add a fast prior predictive checking functionality to \NameOfCode{}. Additionally, giving too many emission lines, without taking into consideration that they may arise from separate structures, or that the combination of emission lines require physical parameters outside of the inference range, can lead to \NameOfCode{} being unable to find physical parameters that match all lines. Because of this, we advise caution on using a long list of emission lines without first studying if those lines come from regions within the object that have similar physical parameters, as this tool will assume they all come from one structure with one set of physical parameters. When taking these factors into consideration, \NameOfCode{} can be a great asset in inferring physical parameter ranges in which to start modeling astrochemical environments.
    
\section{Acknowledgement}
    This project has received funding from the European Union’s Horizon 2020 research and innovation programme under the Marie Skłodowska-Curie grant agreement No 811312 for the project "Astro-Chemical Origins” (ACO)  as well as from the European Research Council (ERC) under the European Union’s Horizon 2020 research and innovation programme MOPPEX 833460.


\bibliography{Bibfile}{}
\bibliographystyle{aasjournal}



\end{document}